\documentclass[twocolumn,prl,citeautoscript]{revtex4-1}

\usepackage{amsmath}
\usepackage{amssymb}
\usepackage{graphicx}
\usepackage{hyperref}
\usepackage{color}

\begin{document}

\title{First-principles prediction of potentials and space-charge layers in all-solid-state batteries}

\author{Michael W. Swift}
\email{swiftmi2@egr.msu.edu}
\affiliation{Department of Chemical Engineering and Materials Science, Michigan State University, East Lansing, Michigan 48824-1226, USA}

\author{Yue Qi}
\email{yueqi@egr.msu.edu}
\affiliation{Department of Chemical Engineering and Materials Science, Michigan State University, East Lansing, Michigan 48824-1226, USA}

\begin{abstract}
As all-solid-state batteries (SSBs) develop as an alternative to traditional cells, a thorough theoretical understanding of driving forces behind battery operation is needed. We present a fully first-principles-informed model of potential profiles in SSBs and apply the model to the Li/LiPON/$\text{Li}_x\text{CoO}_2$ system.  The model predicts interfacial potential drops driven by both electron transfer and Li$^+$ space-charge layers that vary with the SSB's state of charge. The results suggest lower electronic ionization potential in the solid electrolyte favors Li$^+$ transport, leading to higher discharge power.
\end{abstract}

\maketitle

All-solid-state batteries (SSBs) have recently attracted widespread attention due to increased safety and energy density~\cite{Zheng18,Fingerle17,Gittleson17,Leung18}.
Additionally, the functionality of SSBs can serve as the basis for artificial synapses in redox memory devices~\cite{Fuller16,Wang16,Yang17,Yang18}.  Solid-state electronic heterojunctions at solid-electrolyte/electrode interfaces play a critical role in determining ionic transport, and therefore performance, in these devices.  Among various interfacial phenomena such as phase change~\cite{Zhu16}, ionic distribution~\cite{Fingerle17,Santhanagopalan14}, and electrostatic potential drop~\cite{Braun15,Luntz15,Schwobel16}, the formation of a ``space-charge layer'' at the electrode/solid-electrolyte interface is often cited as a barrier for lithium ion transport~\cite{Yamamoto10,Haruyama14}.  The driving force behind space-charge-layer formation is the chemical potential difference between contacting materials, which can result in depletion or enrichment of charged defects near the interface~\cite{Maier95}.  However, even such basic questions as the direction of lithium transfer remain controversial in experiments.  For example, for the commonly studied LiCoO$_2$/LiPON interface, Refs.~\cite{Fingerle17,Gittleson17} reported Li depletion from LiCoO$_2$ to LiPON, while Ref.~\cite{Leung18} reported Li transfer from LiPON to $\text{Li}_x\text{CoO}_2$.  Explicit first-principles calculations of electrode-electrolyte interface structures shed some light on Li transfer and diffusion~\cite{Leung18,Sicolo17}, but it is difficult to draw general conclusions based on a small number of interface structures.

A comprehensive first-principles model is noticeably absent.  Prediction of space-charge layers in solid electrolytes with heterogeneous interfaces has been attempted by lattice models~\cite{Bunde85}, atomistic models~\cite{Morgan11,Stegmaier17}, thermodynamic models~\cite{Fu14,deKlerk18}, and DFT-informed thermodynamic models~\cite{Zhang16}.  However, predicting space-charge-layer formation at the solid-electrolyte/electrode interface is even more challenging, as ion insertion and/or reaction with the electrode alters the material and thus band alignments at the interfaces~\cite{Kasamatsu12}.  In this Letter, we establish an \textit{ab initio} framework to calculate the thermodynamic driving forces and the resulting net interfacial potential drops in a model SSB at equilibrium open-circuit conditions. 
Space-charge layers in the model arise from Li$^+$ transfer (predicted by defect formation energies) and electron transfer (due to interfacial band bending). Together, potential drops and space-charge layers govern the interfacial lithium transport barriers, which are known to be bottlenecks for both performance~\cite{Luntz15} and lifetime~\cite{Lin16}.  Therefore this model represents a fundamental step forward in the theoretical description of SSB interfaces.   

The application of this model to the Li/LiPON/$\text{Li}_x\text{CoO}_2$ system leads to the important discovery that the polarity of the space-charge layer and the sign of the potential drops vary with the state of charge (i.e. Li concentration in $\text{Li}_x\text{CoO}_2$).   This new physics insight unifies the  seemingly contradictory experimental observations~\cite{Fingerle17,Gittleson17,Leung18}.  Improved theoretical understanding also provides valuable design rules for the next generation of devices.  The interfacial electric fields predicted by our model can be engineered to reduce interfacial barriers and cathode overpotential, leading to higher power output.  For the discharge process, favorable interfacial fields may be achieved by restricting operation to lower levels of cathode lithiation or by using solid-electrolyte materials with a high valence band.

The model starts from the assumption of equilibrium open-circuit conditions.  In an SSB, the insulating solid electrolyte blocks the flow of electrons between electrodes, but lithium ions (Li$^+$) are free to move.  As a result, the lithium ion electrochemical potential (the lithium atomic chemical potential minus the electron electrochemical potential) reaches a constant, which is referenced to zero.  
\begin{equation}
	 \tilde\mu_{\text{Li}^+} = \mu_\text{Li}- \tilde\mu_{e^-} = 0\ .
	\label{eq:mu_Li+}
\end{equation}

The difference in $\mu_\text{Li}$ between anode and cathode is the driving force behind battery operation.  This chemical potential difference is the open-circuit voltage (OCV) of the battery, given by the Nernst equation:~\cite{Huggins08}
\begin{equation}
V = (\mu_\text{Li}^a - \mu_\text{Li}^c)/e  \label{eq:Nernst} \ ,
\end{equation}
where superscript $a$ indicates the anode and $c$ the cathode.  The OCV can be calculated as an average for a redox reaction, such as $\text{CoO}_2 + x\,\text{Li} \rightarrow \text{Li}_x\text{CoO}_2$, using the equation  ~\cite{Aykol15,Courtney98} 
\begin{equation}
V = -\left(E[\text{Li}_x\text{CoO}_2] - E[\text{CoO}_2] - x E[\text{Li}]\right)/xe\ ,
\label{eq:OCV}
\end{equation}
where $E$ is the total energy calculated with density functional theory (DFT) and $e$ is the electron charge.

When electrodes are brought into contact through a solid electrolyte, Li$^+$ will flow until the buildup of positive charges in the cathode increases the electrostatic (Galvani) potential $\phi$, blocking further ionic transfer and establishing the open-circuit equilibrium.  This may be seen mathematically by noting that $\mu_\text{Li}= \tilde\mu_{e^-}$ (Eq.~\ref{eq:mu_Li+}) and substituting the definition of electrochemical potential ($\tilde\mu_{e^-} = \mu_{e^-} - e\phi$) into Eq.~\ref{eq:Nernst}: 
\begin{equation}
V = (\mu_{e^-}^a - \mu_{e^-}^c)/e + (\phi^c - \phi^a)  \ . \label{eq:OCVsplit}
\end{equation}
The second term arises from ionic charge accumulation, and is thus called the ``ionic part'' of the OCV: $V_I = \phi^c-\phi^a$~\cite{Gerischer94}.  The first term is the ``electronic part'': $V_{e^-} = (\mu_{e^-}^a - \mu_{e^-}^c)/e$, where $\mu_{e^-}$ is the chemical potential of electrons in the bulk material in absence of any external potential~\cite{Gerischer94}.

The $\mu_{e^-}$ in the SSB may be approximated through the work function $\psi$ with respect to the anode (see [S.III] in the Supplemental Material
~\footnote{See Supplemental Material for details of the computational methodologies used in this work, including Refs.~\cite{Broberg18,Dudarev98,Jain13,Ong15,Ong13,Moses11,VandeWalle87,Freysoldt14,Wang06,Aykol15,Meredig10}}
 for details on this approximation), as  $\mu_{e^-} = -\psi + \psi^a$.   
The ionization potential $IP$  is the position of the valence-band maximum (VBM) below the vacuum level, $\psi$ is the position of the Fermi level below the vacuum, and $E_F$ is the Fermi level defined relative to the VBM.  These quantities are related through $\psi = IP - E_F$.

Therefore Eq.~\ref{eq:mu_Li+} can be rewritten:
\begin{equation}
\tilde\mu_{\text{Li}^+} = \mu_\text{Li} - ( E_F - IP + \psi^a -e \phi ) = 0\ . \label{eq:profile}
\end{equation}

This key equation forms the basis of the potential profile model, which is constructed through the following procedure:

First, a $\mu_\text{Li}$ profile is built.  We choose zero as the $\mu_\text{Li}$ reference in the anode, so by Eq.~\ref{eq:Nernst}, $\mu_\text{Li}=-eV$ in the cathode.  $V$ is calculated using DFT from Eq.~\ref{eq:OCV}.  In the solid electrolyte, $\mu_\text{Li}$ is constrained by thermodynamic stability conditions~[S.I].  The model accepts various assumptions for the $\mu_\text{Li}$ profile in the solid electrolyte [S.II].  A sensible one (which is used in this work) is that $\mu_\text{Li}$ is at its upper limit in the electrolyte near the anode and at its lower limit near the cathode.  
The $\mu_\text{Li}$ profile is completed by interface regions which interpolate $\mu_\text{Li}$ in the electrolyte and electrodes.  For the purposes of this model, the interface is reduced to the net changes that occur in potentials; predictions of the spatial width and microscopic structure of the interface will be left to future work.

Second, the bulk properties of the battery materials are calculated via DFT. To facilitate vacuum alignment, $IP$ is calculated. For metals, $IP = \psi$ [S.III]. For insulators (e.g. solid electrolytes) and semiconductors (e.g. lithium transition metal oxides), point defect calculations that give $E_F$ as a function of $\mu_\text{Li}$ are performed [S.IV].  

Finally, the results of the previous steps are used with Eq.~\ref{eq:profile} to calculate the $\phi$ profile.  Interfacial band bending is caused by bending of the local vacuum level $E_\text{vac}=- e\phi + \psi^a $ (referenced to the anode Fermi level), so a band profile may also be derived from $\phi$.  This completes the potential profile model.

We now apply this general model to the Li/LiPON/$\text{Li}_x\text{CoO}_2$ SSB system.  Our DFT calculations use \textsc{vasp}~\cite{Kresse96} with the projector-augmented wave (PAW) method~\cite{Blochl94}, employing a 520 eV cutoff and the PBE~\cite{Perdew96} functional.  In order to correctly describe layered $\text{Li}_x\text{CoO}_2$, the van der Waals interaction must be taken into account~\cite{Aykol15}.  The optPBE approach~\cite{Dion04,Perez09,Klimes10,Klimes11} is used as recommended in Ref.~\cite{Aykol15}; tests using other ``opt'' functionals produce similar results.  Correlations on the cobalt $d$ states are taken into account through DFT+$U$~\cite{Dudarev98} with $U=3.32$ eV~\cite{Wang06,Aykol15}.  
The calculated OCV is $V = 4.02$ V for LiCoO$_2$ and $V=4.15$ V for $\text{Li}_{0.5}\text{CoO}_2$, consistent with both DFT-predicted and experimental values~\cite{Aykol15,Amatucci96}.

LiPON electrolytes are typically amorphous and have widely varying stoichiometry, but are generally characterized by oxygen-decorated P-N-P chains ionically bonded to lithium.~\cite{Sicolo16}
In order to facilitate first-principles calculations, a crystalline structure containing these structural motifs, \emph{SD}-Li$_2$PO$_2$N, is used.  It has been shown experimentally to have similar lithium transport properties to glassy LiPON~\cite{Senevirathne13,Holzwarth14}.  
The band structure is shown in Fig. S.4. 

\begin{figure}
\includegraphics[width=\columnwidth]{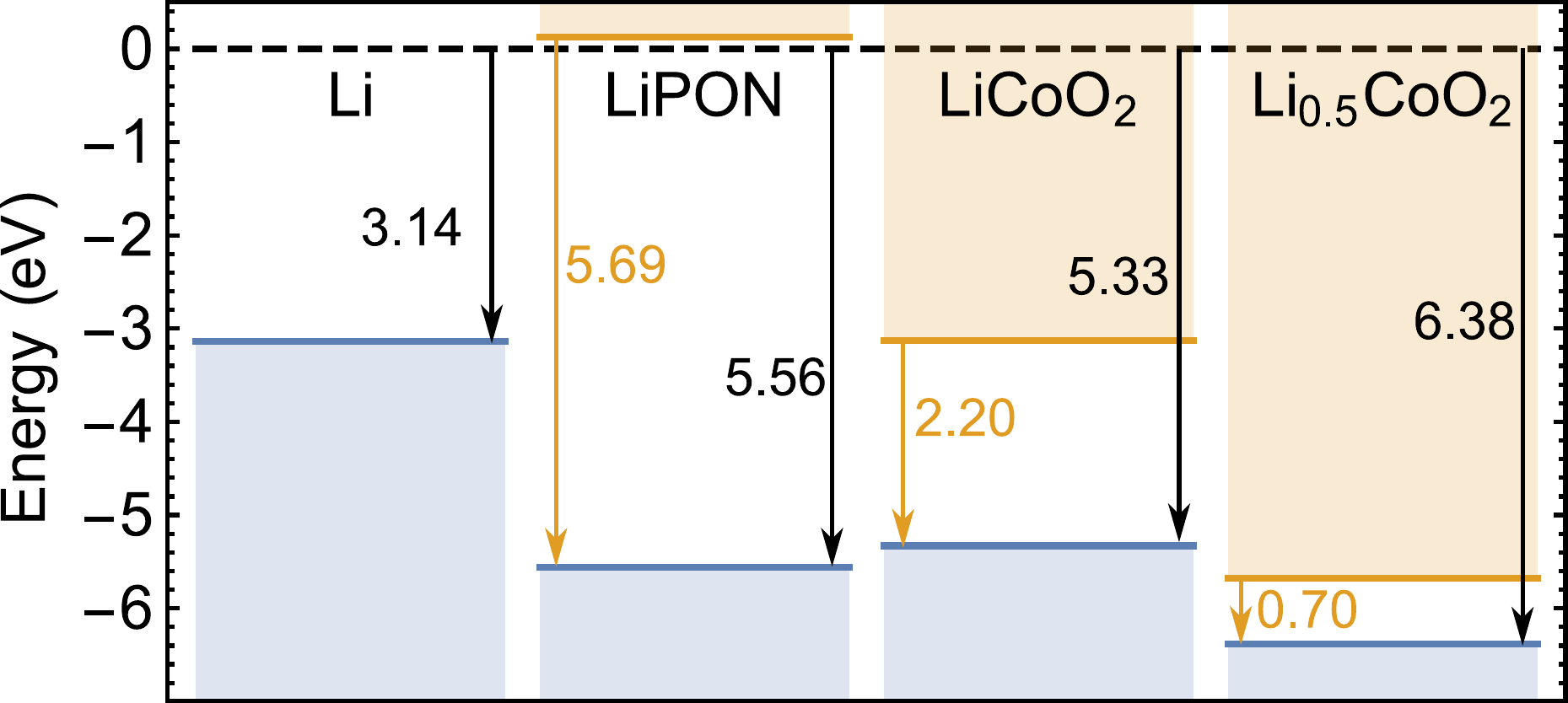}
\caption{The vacuum-aligned position of the electronic bands in lithium metal, LiPON, $\text{LiCoO}_2$, and $\text{Li}_{0.5}\text{CoO}_2$ before contact.  
Occupied bands (VBM) are blue and unoccupied bands (CBM) are orange.  Vacuum (dashed line) alignments and band gaps are indicated.  }
\label{fig:bands}
\end{figure}

Vacuum alignments are calculated using standard procedures~[S.III].  Calculated band gaps and ionization potential for SSB materials before contact are shown in Fig.~\ref{fig:bands}, and $IP$ and $\mu_\text{Li} = \tilde\mu_{e^-}$ are tabulated in Table~\ref{table}.

\setlength{\tabcolsep}{5pt}
\begin{table}
\begin{tabular}{lrrrrr}
                        & $E_F$        & $IP$         & $\tilde\mu_{e^-}$ & $V_{e^-}$ & $V_I$\\
Li metal                & 0            & 3.14         & 0                 \\
LiPON (Li-rich)         & 2.02         & 5.56         & $-0.68$           \\
LiPON (Li-poor)         & 1.02         & 5.56         & $-2.64$           \\
LiCoO$_2$               & 0            & 5.33         & $-4.02$           & 2.19      & 1.82 \\
Li$_{0.5}$CoO$_2$       & 0            & 6.38         & $-4.15$           & 3.24      & 0.91 \\
\hline
                        & $\Delta E_B$ & $\Delta E_I$ & $\Delta\phi$    \\
Li/LiPON                & $-2.70$      & $-2.42$      & $+0.28$         \\
LiPON/LiCoO$_2$         & $-0.37$      & $+0.23$      & $+0.59$         \\
LiPON/Li$_{0.5}$CoO$_2$ & $-0.49$      & $-0.82$      & $-0.33$     
\end{tabular}
\caption{Top: bulk quantities relevant to Eq.~\ref{eq:profile} for SSB materials, calculated using DFT.  Electronic and ionic parts of OCV (Eq.~\ref{eq:OCVsplit}) with respect to Li metal are shown for Li$_x$CoO$_2$. Bottom: band offsets and electrostatic potential drops for SSB interfaces (Eq.~\ref{eq:deltaphi}). }
\label{table}
\end{table}

In order to determine the direction of ion transfer at interfaces and the Fermi level inside LiPON, the formation energies of point defects must be calculated and compared for both LiPON and $\text{Li}_x\text{CoO}_2$. The formation energy of point defect $X$ in charge state $q$ is given by~\cite{Freysoldt14}~[S.IV]
\begin{equation}
E^f[X^q] = E_\text{tot}[X^q] - E_\text{tot}[\text{bulk}] - \sum_i n_i \mu_i + qE_F.
\label{eq:Ef}
\end{equation}
The $\mu_\text{Li}$ has already been discussed in detail.  Other atomic chemical potentials are set by thermodynamic stability~[S.I].  Supercell artifacts from charged defect calculations are corrected using the Freysoldt method~\cite{Freysoldt09,Freysoldt11}.  Defect formation energy calculations are partially automated using pymatgen~\cite{Ong13} and PyCDT~\cite{Broberg18}.  

A thorough inventory of point defects in LiCoO$_2$~[S.IV, Fig.~S.2] shows that, at the cathode chemical potential $\mu_\text{Li} = -4.02$ eV, the dominant defect is the shallow acceptor $V_\text{Li}^-$.  As a result, the material is heavily $p$-type, in agreement with experiments~\cite{Tukamoto97}, so $E_F=0$.  The Fermi level in $\text{Li}_{0.5}\text{CoO}_2$ is similarly at the top of the valence band.

\begin{figure}
\includegraphics[width=\columnwidth]{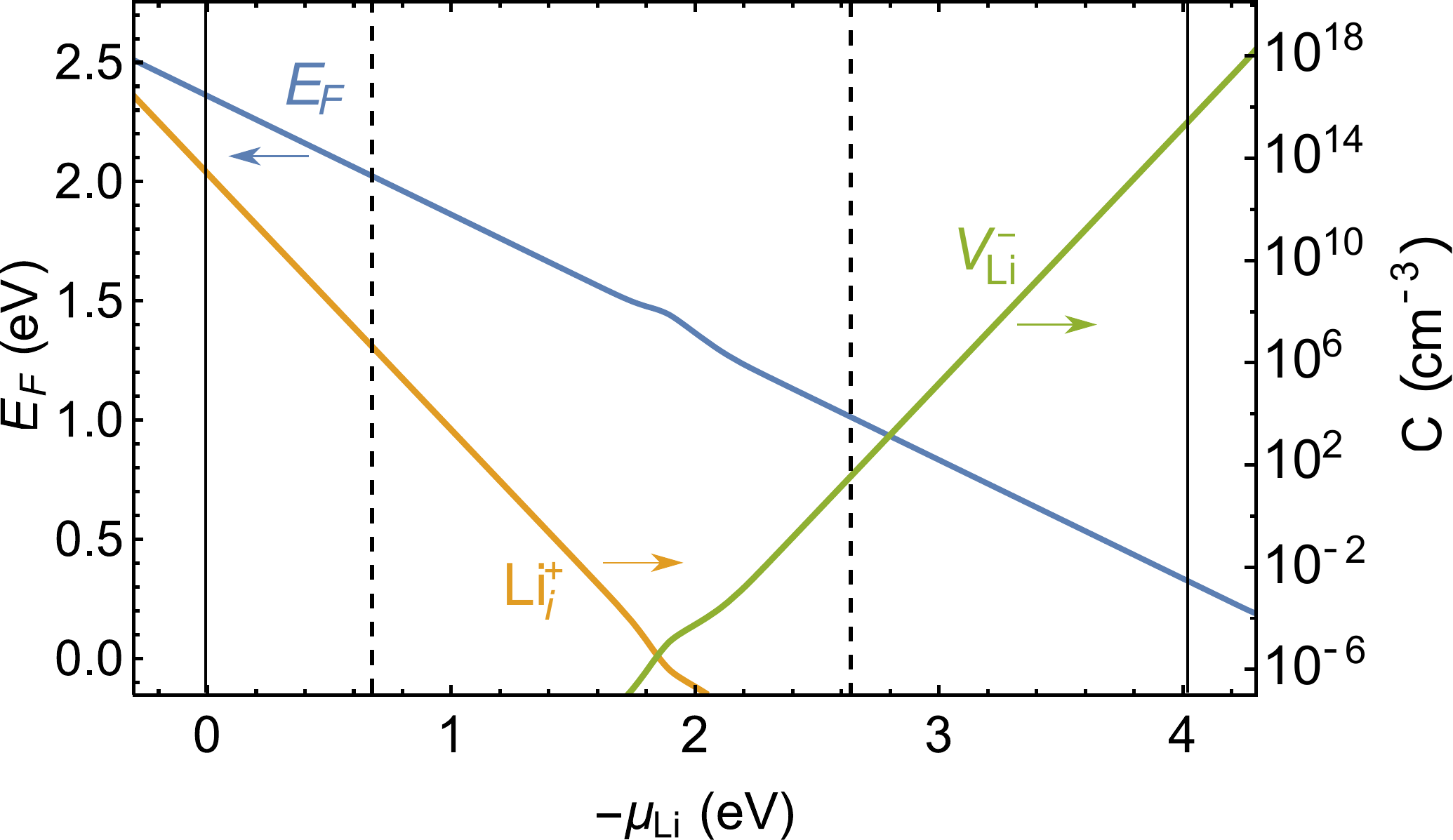}
\caption{Fermi level referenced to the VBM ($E_F$, blue; left axis) and Li$^+$ carrier defect concentrations (Li$_i^+$, orange and $V_\text{Li}^-$, blue; right axis) as a function of $\mu_\text{Li}$.  Li metal and LiCoO$_2$ values for $\mu_\text{Li}$ are solid vertical lines, and the boundaries for thermodynamic stability of LiPON are dashed lines. }
\label{fig:LiPON_muLi}
\end{figure}

Since the LiPON electrolyte exists in between the Li-rich and Li-poor extremes of the electrodes, the defect formation energies are considered as a function of $\mu_\text{Li}$.  This allows calculation of the $E_F$ as a function of $\mu_\text{Li}$ through the requirement of charge neutrality.  The results are shown in Fig.~\ref{fig:LiPON_muLi}, with details in~[S.IV, Fig. S.3].  In the dilute limit, the concentration of a defect is given by $c=N e^{-E^f/k_B T}$ where $N$ is the concentration of defect sites and $k_B T$ is the temperature in energy units.  This allows for the concentration of Li$^+$ carriers to be calculated as a function of $\mu_\text{Li}$ (Fig.~\ref{fig:LiPON_muLi}).  Note that the dominant lithium carrier in LiPON changes from interstitials in anode-like conditions to vacancies in cathode-like conditions.

Defect formation energies also determine the direction of Li$^+$ transfer at interfaces.  At the Li/LiPON interface, Li$^+$ is the dominant carrier in LiPON.  The energy change of the reaction $\text{Li}^+_i[\text{Li}] \rightarrow \text{Li}^+_i[\text{LiPON}] $ determines the direction of Li transfer:
\begin{equation}
E^f[\text{Li}^+_i; \text{LiPON}] - E^f[\text{Li}^+_i; \text{Li}] = - 0.17 \text{ eV}. \label{eq:Li-LIPON}
\end{equation}
This means that Li$_i^+$ will tend to move from lithium metal into LiPON.  

At the LiPON/LiCoO$_2$ interface, the dominant lithium carrier is $V_\text{Li}^-$, so the relevant reaction is $V_\text{Li}^-[\text{LiPON}] \rightarrow V_\text{Li}^-[\text{LiCoO$_2$}]$. The reaction energy is $- 0.28$ eV, so $V_\text{Li}^-$ moves from LiPON into LiCoO$_2$.
For $x=0.5$, the reaction energy is 8.52 eV.  This represents a very strong tendency for $V_\text{Li}^-$ to migrate from $\text{Li}_{0.5}\text{CoO}_2$ into LiPON.
These results indicate the direction of Li flow at the interface depends on the Li concentration in $\text{Li}_x\text{CoO}_2$, i.e. the battery's state of charge (SOC). 

In addition to Li$^+$, transfer of electrons or holes may contribute to interface dipoles.  The net charge transfer is caputred by the electrostatic potential drop $\Delta \phi$ at each interface, which causes electronic band bending.  $\Delta \phi$ may be extracted from the full potential profile that will be calculated using Eq.~\ref{eq:profile}, but it is worthwhile to examine $\Delta\phi$ directly in the context of an interface.  Eq.~\ref{eq:profile} shows that the electrostatic potential drop is given by the changes in bulk quantities across the interface: 
\begin{equation}
	\Delta\phi = \frac{1}{e}\left(\Delta E_F - \Delta IP - \Delta \tilde \mu_{e^-}\right) \ .\label{eq:deltaphi}
\end{equation}

This equation may be recast in terms analogous to traditional band-bending theory at semiconductor heterojunctions~\cite{Sze12}, with the key difference that it is not $E_F$, but rather $\tilde\mu_{\text{Li}^+}$, that attains a constant value in equilibrium.  The intrinsic valence band offset at the interface is $\Delta E_I = -\Delta IP$ (a positive value indicates a higher valence band in the second material).  
The difference in bulk VBM positions is $\Delta E_B = \Delta \tilde \mu_{e^-} - \Delta E_F$.  The  difference between $\Delta E_I$ and $\Delta E_B$ is caused by band bending, driven by the interfacial potential drop $\Delta \phi = (\Delta E_I - \Delta E_B)/e$.

\begin{figure}
\includegraphics[width=\columnwidth]{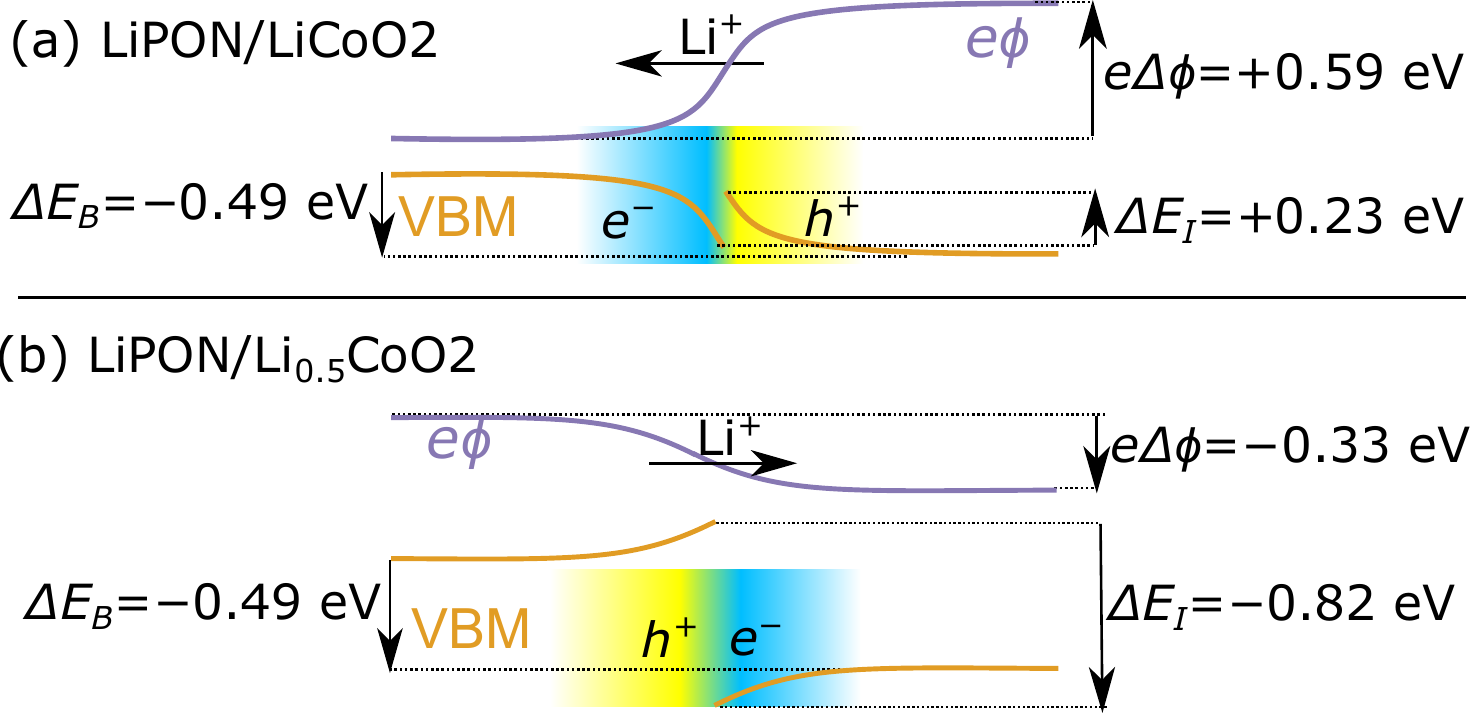}
\caption{Schematic of the electrostatic potential and valence bands at the interfaces between LiPON and $\text{Li}_x\text{CoO}_2$ for (a) $x=1$ and (b) $x=0.5$.  Electron transfer based on the band structure and Li$^+$ transfer based on defect formation energies are illustrated schematically.  The shapes of the curves are for illustrative purposes only.  Computed bulk offsets $\Delta E_B$, intrinsic offsets $\Delta E_I$, and resulting potential shift $e\Delta\phi$ are indicated.  }
\label{fig:interface}
\end{figure}

Based on $E_F$, $IP$, and $\mu_\text{Li}$ on either side of interfaces, the  $\Delta E_I$, $\Delta E_B$, and $\Delta \phi$ at three interfaces are calculated, as shown in Table~\ref{table}.
At the Li/LiPON interface, the increase in $\Delta \phi$ is mainly caused by Li$^+$ transferred to LiPON, since no electron transfer into LiPON is possible because the Fermi level is deep in the forbidden gap. For the cathode interfaces, schematics of band-bending and charge transfer are shown in Fig.~\ref{fig:interface}.  At the LiPON/LiCoO$_2$ interface, a positive $\Delta\phi$ predicts net transfer of positive charge into LiCoO$_2$: electrons are transferred into LiPON, outnumbering the Li$^+$ that move into LiPON.  Fingerle \textit{et al.}~have measured the work function of LiCoO$_2$ and found it bent 0.3 eV up toward LiPON at this interface~\cite{Fingerle17}, consistent with our prediction.  At the LiPON/Li$_{0.5}$CoO$_2$ interface, $\Delta\phi$ is negative, so net positive charge is transferred into LiPON.  Electrons are transferred out of LiPON, again outnumbering the Li$^+$ that move into the cathode.  This shows that the potential drop at the cathode interface is driven primarily by electron transfer.


\begin{figure}
\includegraphics[width=\columnwidth]{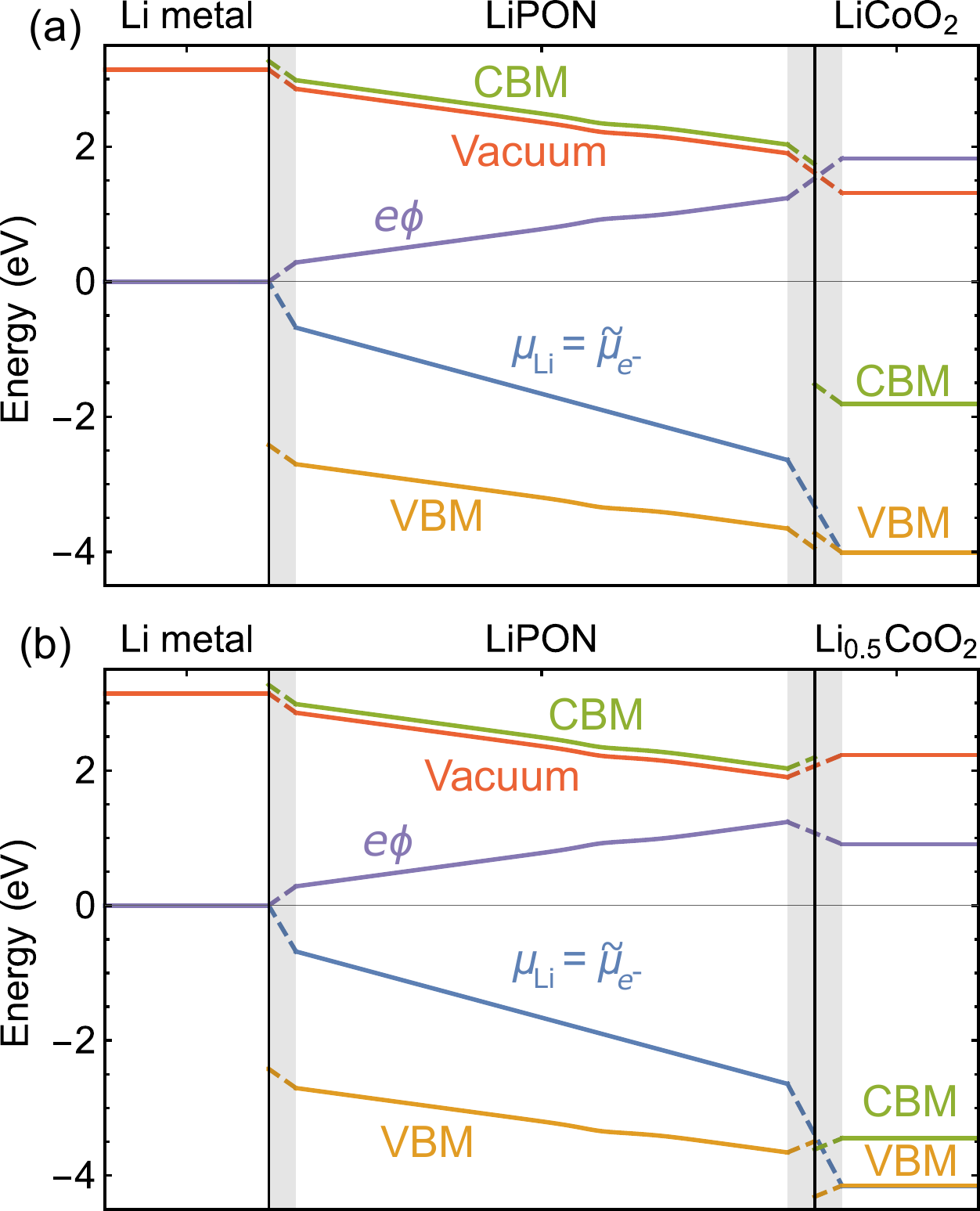}
\caption{Thermodynamic potentials and band edges in a model 1D Li/LiPON/$\text{Li}_x\text{CoO}_2$ battery in open circuit conditions.  (a) Discharged state ($x=1$)  (b) Charged state ($x=0.5$) }
\label{fig:profile}
\end{figure}

The full Li/LiPON/$\text{Li}_x\text{CoO}_2$ profiles calculated from Eq.~\ref{eq:profile} and synthesizing all this information are shown in Fig.~\ref{fig:profile}.  
The interface regions on both anode and cathode side are indicated in gray, and the calculated quantities on either side of a given interface are connected by dotted lines, showing the net change across the interface.  While the details of the potential profile across the interfaces are unknown (and may depend on complex kinetics), the net change is enough to draw useful conclusions about the effects of the interfaces on battery operation.  In the future, this framework can easily be extended to explicitly include SEI phases or interlayers, providing more detail in the interfacial region.

Our results explain the different directions of Li$^+$ transfer observed at the LiPON/$\text{Li}_x\text{CoO}_2$ interface.  Experiments showing Li$^+$ transfer into $\text{Li}_x\text{CoO}_2$~\cite{Leung18} may have had lower lithiation than experiments which observe Li$^+$ transfer into LiPON~\cite{Fingerle17,Gittleson17}.  

The direction of Li$^+$ transfer and the interface dipole in SSBs have important implications for device performance~\cite{Leung15}.  An electric field at the interface which attracts Li$^+$ into the cathode is desirable since it reduces the energy barrier for discharge. When electrons are allowed to flow through an external circuit, a lower interfacial barrier will lead to a reduced overpotential and higher discharge power output.  
Our results show that the Li/LiPON interface always adds a barrier for discharge and the LiPON/cathode interface potential drop will reduce the discharge barrier at high SOC and increase it at low SOC.  Linear interpolation predicts the unfavorable dipole develops for lithiation greater than $x=0.68$.  This effect is compounded by changes in lithium carrier concentration within the electrolyte.  Since the lithium carrier in LiPON near the cathode is $V_\text{Li}^-$, lithium transfer into the LiPON at full cathode lithiation reduces the carrier concentration, further hindering the discharge process.  By contrast, Li$^+$ transfer into $\text{Li}_{0.5}\text{CoO}_2$ increases the concentration of vacancies, assisting discharge.  This prediction agrees with measured impedance changes upon Li$^+$ transfer~\cite{Gittleson17}. Our results suggest that the changes they discuss in Li$^+$ concentration profiles are only part of the story: the interfacial potential drop also plays a crucial role.  In the opposite direction, the potential drop at the LiPON/cathode interface at high SOC will resist Li$^+$  transport for fast charging.  

Based on this model, several methods can be suggested to reduce the interface resistance of SSBs during discharging.  One approach is raising the electrolyte valence band.  This tends to lower $\Delta\phi$, maintaining a favorable interface dipole for discharge across a wider range of lithiation and thus increasing power output without sacrificing capacity.  This could be done by tuning LiPON growth methods~\cite{Hausbrand15} or by using alternative electrolyte materials that maintain a negative band offset with $\text{Li}_x\text{CoO}_2$ across a wider range of lithiation.  Alternatively, higher $\Delta\phi$ is desired for fast charging, suggesting the opposite approach depending upon the desired outcome.
Another approach would be to apply an interlayer between the solid electrolyte and the electrode to mitigate the interfacial barrier. 
This may occur in the gray areas in Fig.~\ref{fig:profile}, where LiPON is not thermodynamically stable and may form an SEI layer. Our model can be extended to explicitly model the role of such interlayers in modifying the potentials and carrier concentration at electrolyte/interlayer/electrode interfaces~\cite{Gittleson17}.

In conclusion, we have presented a new technique for building a potential profile in a model SSBs based on inputs from first-principles calculations. 
This model predicts various key properties of the interfaces between electrodes and electrolyte.  At the $\text{Li}_x\text{CoO}_2$ interface, our results suggest that the interfacial potential drop is driven by electron transfer, and the direction of $e^-$ and Li$^+$ transfer depends on the degree of lithiation of the cathode (equivalently the SOC of the battery).  These results unify the conflicting Li-transfer trends observed in experiments, and suggest design rules for improving power output by minimizing discharge barriers at the electrode/electrolyte interfaces.  The methodology developed in this work is broadly applicable to modeling other all-solid-state battery systems.

\begin{acknowledgments}
Thanks to K. Leung at Sandia and A. Van der Ven at UCSB for helpful suggestions. This work was supported by the
Nanostructures for Electrical Energy Storage (NEES) center, an Energy
Frontier Research Center funded by the U.S. Department of Energy,
Office of Science, Basic Energy Sciences under Award number
DESC0001160.  
This work was supported in part by Michigan State University through computational resources provided by the Institute for Cyber-Enabled Research.
\end{acknowledgments}

\bibliography{mybib}

\end{document}